# Search for Exotic Mesons in $\pi^- p$ Interactions at 18 GeV/c*

J. M. LoSecco, T. Adams, J. M. Bishop, N. M. Cason, J. J. Manak,
A. H. Sanjari, W. D. Shephard, D. L. Stienike, S. A. Taegar, D. R. Thompson
*Department of Physics, University of Notre Dame,*
*Notre Dame, IN 46556, USA*

S. U. Chung, R. W. Hackenburg, C. Olchanski, D. P. Weygand, H. J. Willutzki
*Department of Physics, Brookhaven National Laboratory,*
*Upton, L.I.,NY 11973, USA*

S. Denisov, A. Dushkin, V. Kochetkov, I. Shein, A. Soldatov
*Institute for High Energy Physics,*
*Protvino, Russian Federation*

B. B. Brabson, R. R. Crittenden, A. R. Dzierba, J. Gunter,
R. Lindenbusch, D. R. Rust, E. Scott, P. T. Smith, T. Sulanke, S. Teige
*Department of Physics, Indiana University,*
*Bloomington, IN 47405, USA*

Z. Bar-Yam, J. P. Dowd, P. Eugenio, M. Hayek, W. Kern, E. King
*Department of Physics, University of Massachusetts Dartmouth,*
*North Dartmouth, MA 02747, USA*

V. A. Bodyagin, A. M. Gribushin, O. L. Kodolova, M. A. Kostin, V. L. Korotkikh,
A. I. Ostrovidov, A. S. Proskuryakov, L. I. Sarycheva, N. B. Sinev,
I. N. Vardanyan, A. A. Yershov
*Institute for Nuclear Physics, Moscow State University,*
*Moscow, Russian Federation*

D. S. Brown, T. K. Pedlar, K. K. Seth, J. Wise, D. Zhao
*Department of Physics, Northwestern University,*
*Evanston, IL 60208, USA*

G. S. Adams, J. Napolitano, M. Nozar, J. A. Smith, M. Witkowski
*Department of Physics, Rensselaer Polytechnic Institute,*
*Troy, NY 12180, USA*

*Presented at the International School of Nuclear Physics, Erice, Sicily, Italy, September 1995

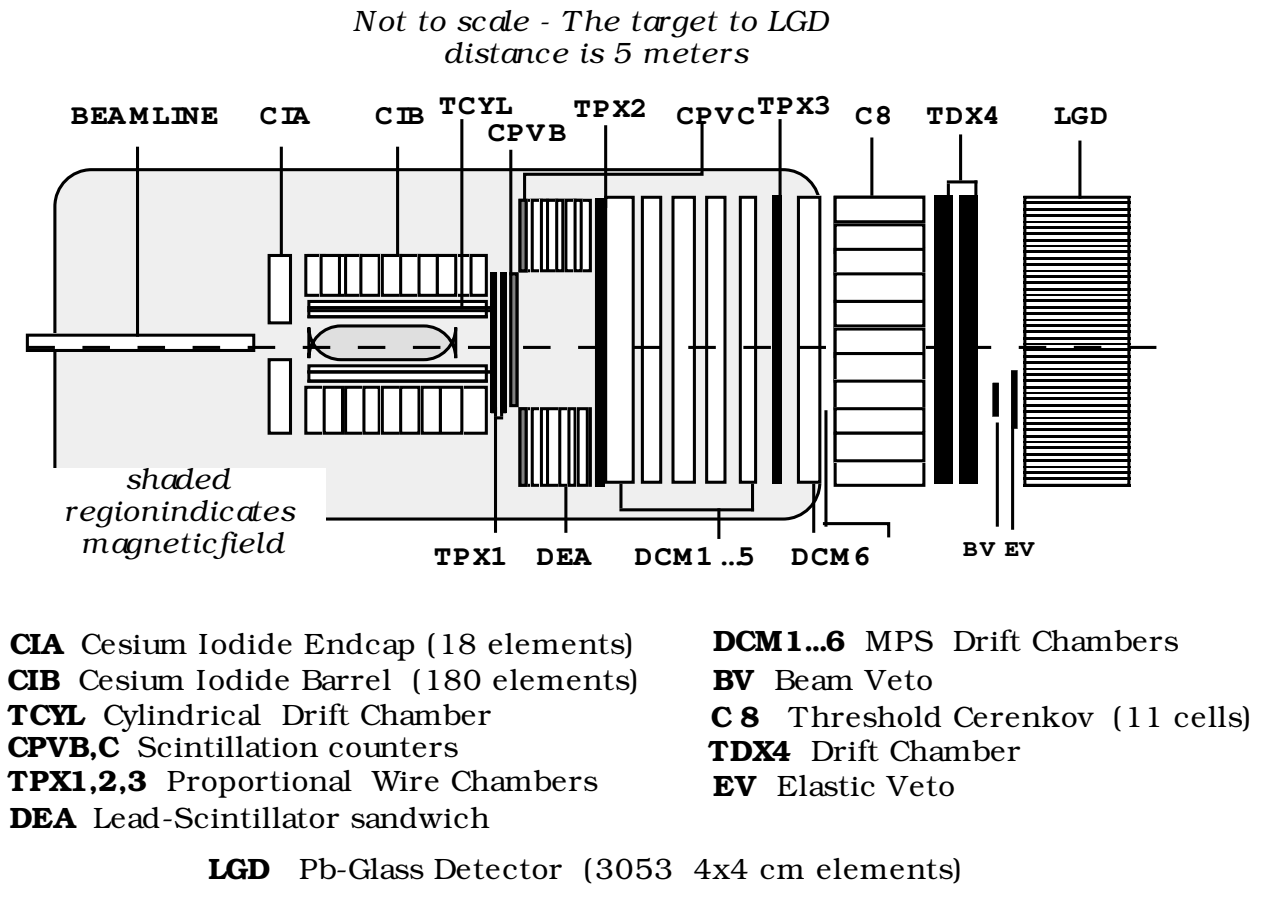

Figure 1: Schematic View of the Detector

# Introduction

Quantum Chromodynamics (QCD) is a theory of interacting quarks and gluons. The distinctive charge and color properties are responsible for the quantum numbers of most of the known hadrons. Exotic mesons are by definition mesons that are not composed of a quark antiquark pair. Some of these exotic states can be identified by the fact that they have quantum numbers that can never be formed from a quark antiquark pair. In particular the quantum numbers $J^{PC}$= $0^{--}$, $0^{+-}$, $1^{-+}$, $2^{+-}$, $3^{-+}$ *etc.* are explicitly exotic. Observation of states with these quantum numbers would be direct evidence for physics beyond the naive quark model. Such states can easily be formed in QCD as a quark antiquark gluon bound state. Measurement of the mass, width and branching ratios of $q\bar{q}g$ exotics will give us a great deal of information about the role of the gluon in QCD.

# Experiment

Bag model and lattice QCD estimates imply that exotics should exist in the same mass region as ordinary hadrons. This complicates the search for exotics by requiring that the experiment deal with large numbers of conventional states. On the other hand the presence of known states provides the opportunity to observe the interference produced by the decay of different states to the same final state.

Some guidance as to what final states may be populated by the decay of exotic mesons is suggested by the "Flux Tube Model"[1]. It is suggested that mesons with $L = 1$ are preferred in such decays so that searches in the final states $a_0\pi$, $f_1\pi$, $a_2\pi$, $b_1\pi$ are more likely to be fruitful than more conventional multi meson final states. In addition a number of groups have reported evidence for $1^{-+}$ hybrids in the $\eta\pi$[2] and $f_1\pi$[3] final states.

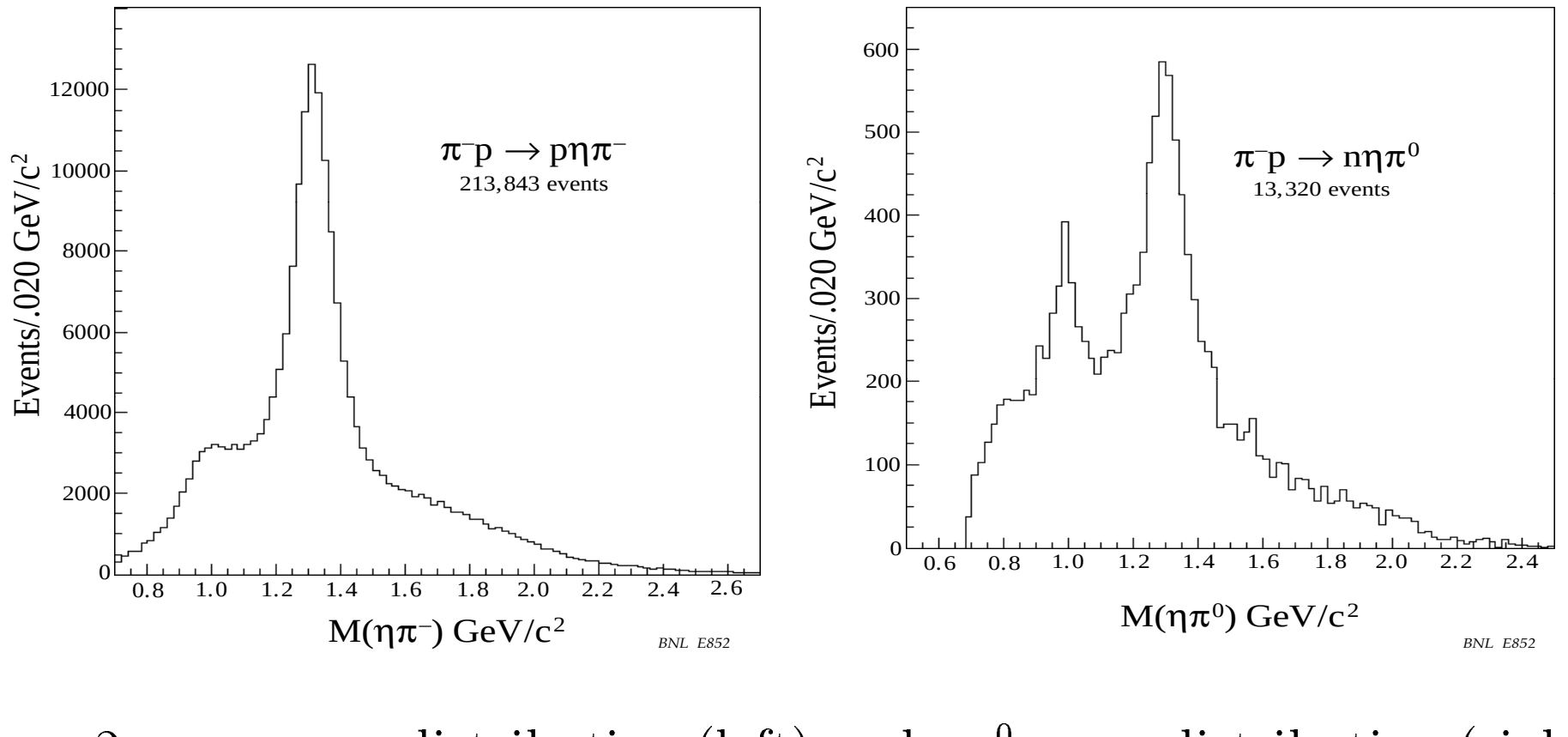

Figure 2: $\eta\pi^-$ mass distribution (left) and $\eta\pi^0$ mass distribution (right).

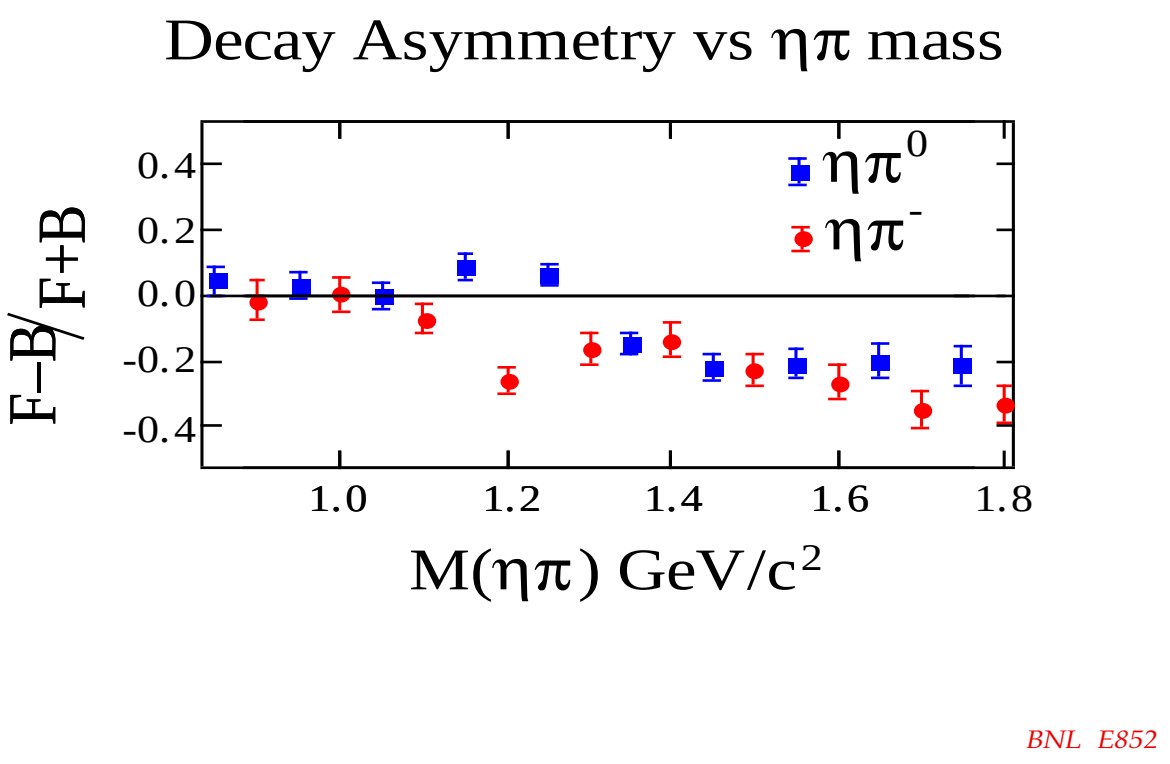

Figure 3: Forward-Backward Asymmetry for $\eta\pi$. The boxes are for $\eta\pi^0$ and the circles are for $\eta\pi^-$

Studies of radiative $J/\psi$ decay have also indicated that the $\eta$ and $\eta'$ mesons have a large "glue" component and would be prevalent in the decay of states with an explicit gluon content.

Brookhaven experiment 852 has been designed to search for exotics in $\pi^- p$ reactions at 18 Gev/c. The detector is shown in figure 1. Electromagnetic decays are detected in a 3049 element lead glass calorimeter located 5 meters downstream of a 30 cm liquid hydrogen target. Charged tracks are detected and measured with six drift chambers each of which contains 7 wire planes. The tracking region has a 1 Tesla magnetic field. Triggering is done with proportional wire chambers that measure charged multiplicity, and with a hardwired processor that is triggered on mass, energy or multiplicity in the calorimeter.

## $\eta\pi$

Some evidence for an exotic $1^{-+}$ state has been found in the $\eta\pi^0$ final state[2]. We have looked at the $\eta\pi^0$, $\eta\pi^-$ and $\eta'\pi^-$ final state. The first two of these (figure 2) show strong evidence for the $2^{++}$ state, the $a_2(1320)$, which is an isovector.

Interference of the $a_2(1320)$ with a $1^{-+}$ amplitude will produce an asymmetry. Such an asymmetry has been observed. The asymmetry as a function of mass is plotted in figure 3. The asymmetry is indicative of interference between odd and even waves. It is a necessary, but not sufficient condition for the presence of an exotic $1^{-+}$ resonance. The existence of a resonance must be determined by a detailed partial wave analysis to study the phase variation of the $1^{-+}$ amplitude. The phase is used to distinguish a resonance from a non resonant $\eta\pi$ P wave. Work on the partial wave analysis is still in progress.

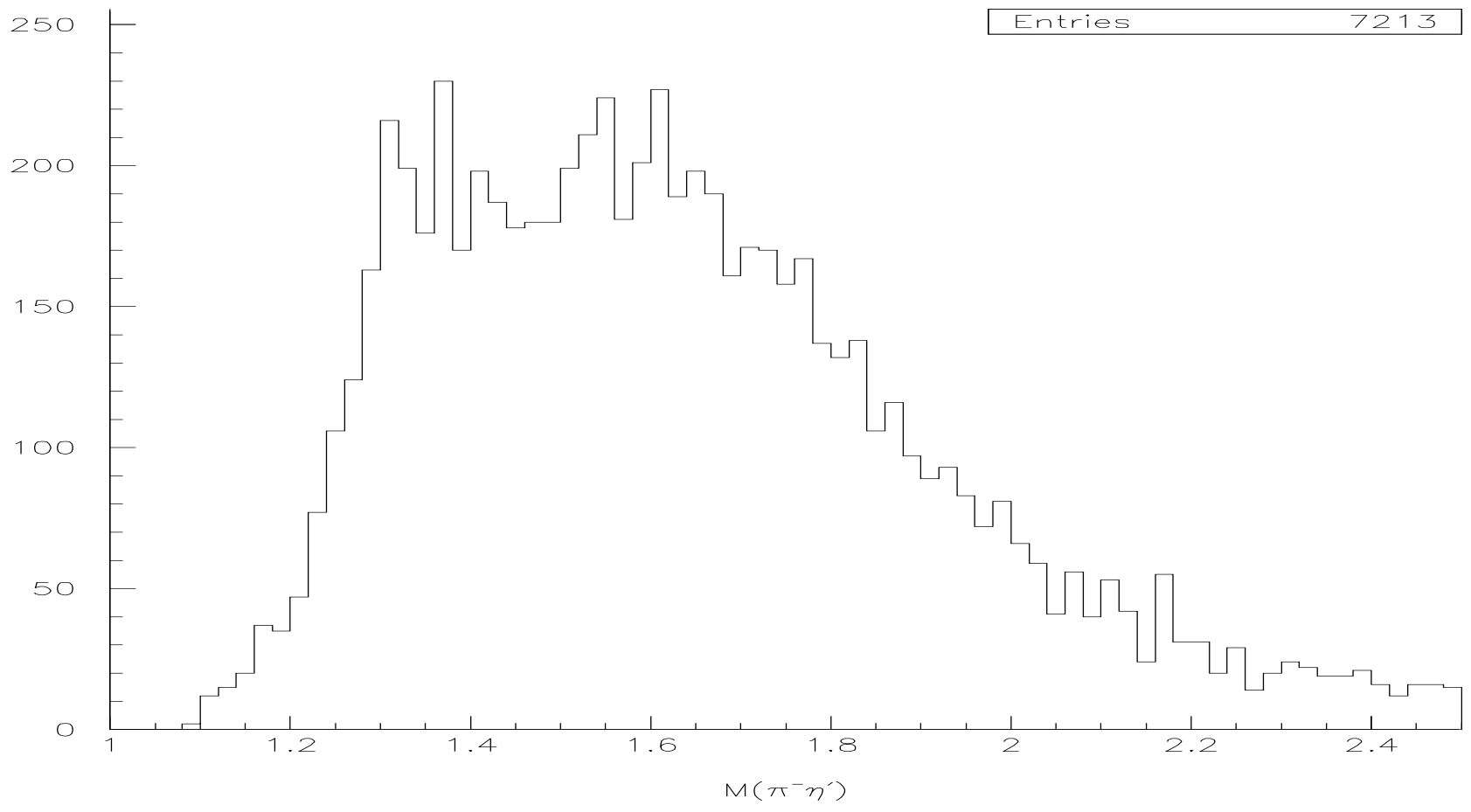

Figure 4: $\eta'\pi^-$ mass plot

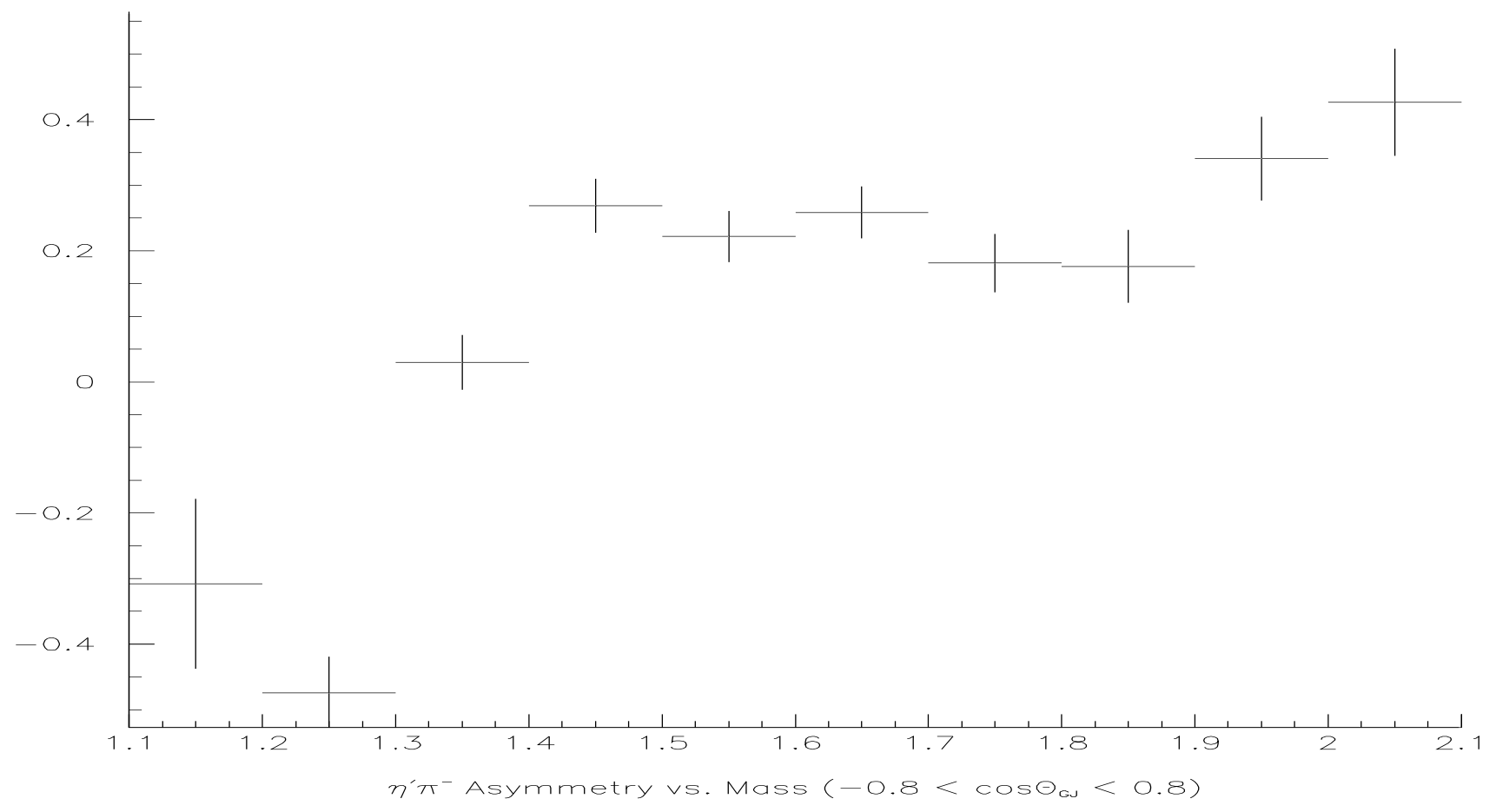

Figure 5: $\eta'\pi^-$ asymmetry as a function of mass

# $\eta'\pi^-$

The VES experiment has reported [4] significant $1^{-+}$ amplitude in the $\eta'\pi^-$ final state. We observe this channel via the decay $\eta' \to \eta\pi^+\pi^-$ (44%). Figure 4 shows the mass distribution for $\eta'\pi^-$. There is some evidence for the $2^{++}$ state the $a_2(1320)$.

Figure 5 shows a plot of the asymmetry as a function of mass of the $\eta'\pi^-$ system. The asymmetry again indicates the presence of $1^{-+}$. Preliminary results from a partial wave analysis are in good agreement with VES [4].

# $f_1\pi$

The $f_1\pi$ state has been identified with a $1^{-+}$ exotics candidate [3] in the 1.6-2.2 GeV/c$^2$ mass range. Figure 6 shows the $\eta\pi^+\pi^-$ mass distribution in the $\eta\pi^+\pi^-\pi^-$ data sample. The $\eta'(958)$ and the $f_1(1285)$ are clearly visible. Figure 7 shows the $f_1\pi$ mass distribution. This state is studied via the decay chain $f_1 \to a_0\pi$ and $a_0 \to \eta\pi$.

# $\pi^+\pi^-\pi^0\pi^0$

The $\pi^+\pi^-\pi^0\pi^0$ state is potentially interesting since it gives access to the preferred decay mode $a_2\pi$ (figure 8) via the decay sequence $a_2 \to \rho\pi$ (70%) and $\rho \to \pi\pi$.

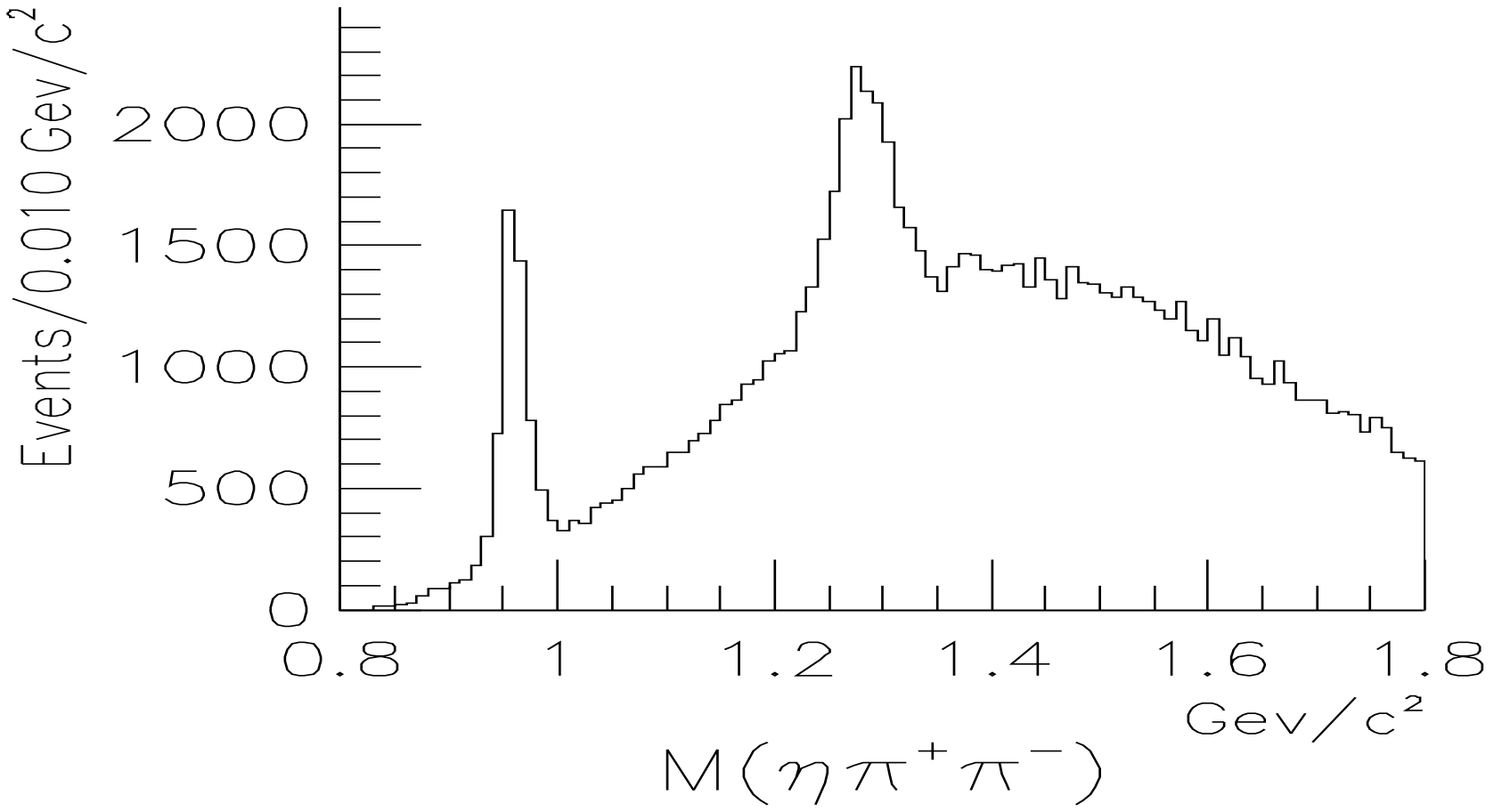

Figure 6: $\eta\pi^+\pi^-$ mass plot from the $\eta3\pi$ sample. Note the $\eta'(958)$ and $f_1(1285)$ peaks.

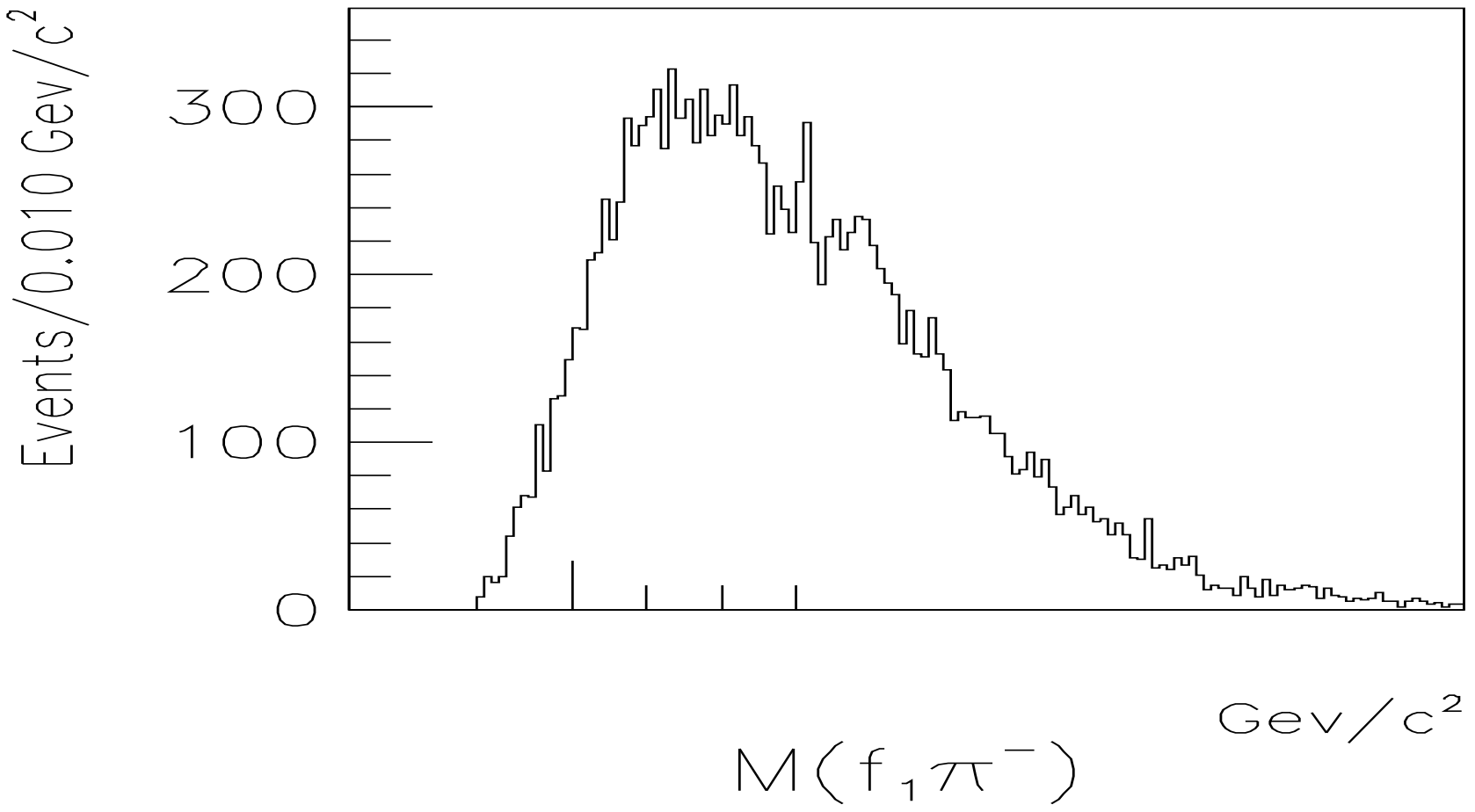

Figure 7: $f_1\pi^-$ mass plot

In addition the final states $\rho^+\rho^-$ (figure 9) and $\omega\pi^0$ will populate this channel. $\omega\pi^0$ is shown in figure 10. The $b_1$ and $\rho_3(1690)$ are clearly visible.

# $b_1\pi$

The $b_1\pi$ states are preferred decay modes for hybrids in a number of models[1]. The $b_1$ and the $\pi$ are both isovector particles so the $b_1\pi$ decay mode can be accessed by both isovector and isosinglet states. The $b_1(1235)$ decays almost entirely via $\omega\pi$. The $\omega$ has a large branching ratio (89%) into $\pi^+\pi^-\pi^0$. So one would expect a strong signal for $b_1\pi$ in $\pi^+\pi^-\pi^0\pi^-\pi^0$. We note a strong $b_1$ signal in the $\pi^+\pi^-\pi^0\pi^0$ as shown in figure 10. In addition $\pi^+\pi^-\pi^0\pi^-\pi^0$ is coupled to $\omega\rho$.

Figure 11 shows the $\omega$ mass region of the $\pi^+\pi^-\pi^0$ mass plot in the 5 $\pi$ sample. A very clear $\omega$ is visible. This $\omega$ can be combined with either of the two remaining $\pi$'s to look for a $b_1^-$ or a $b_1^0$. Figure 12 shows both of these $\omega\pi$ plots. Neither has much of an enhancement in the vicinity of the $b_1(1235)$. The absence of a clear $b_1$ signal makes it less likely that a strong signal will be found in the preferred $b_1\pi$ final state.

Figure 13 shows the $\omega\pi\pi$ mass plot. The $a_2(1320)$ is clearly visible here.

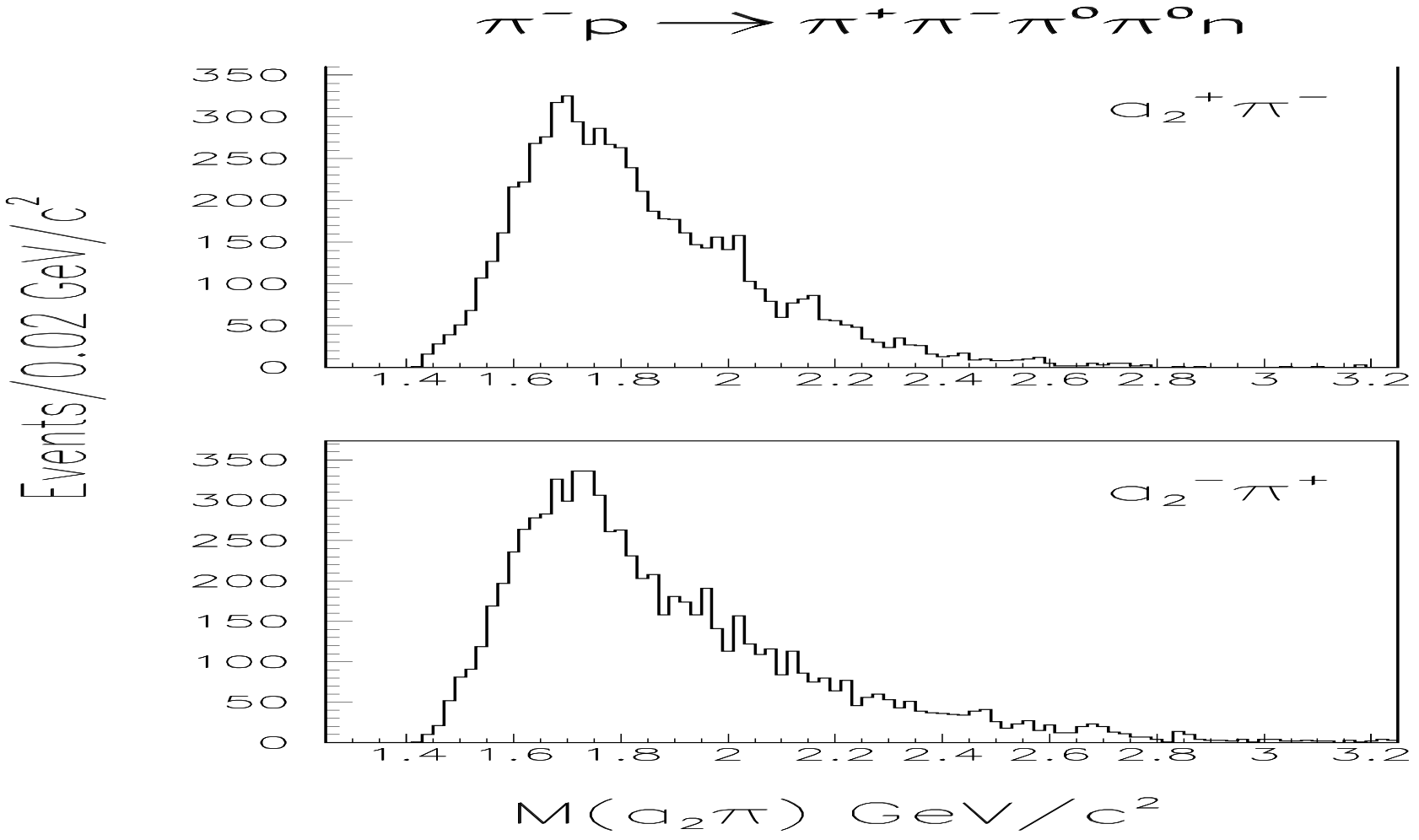

Figure 8: $a_2\pi$ mass plots.

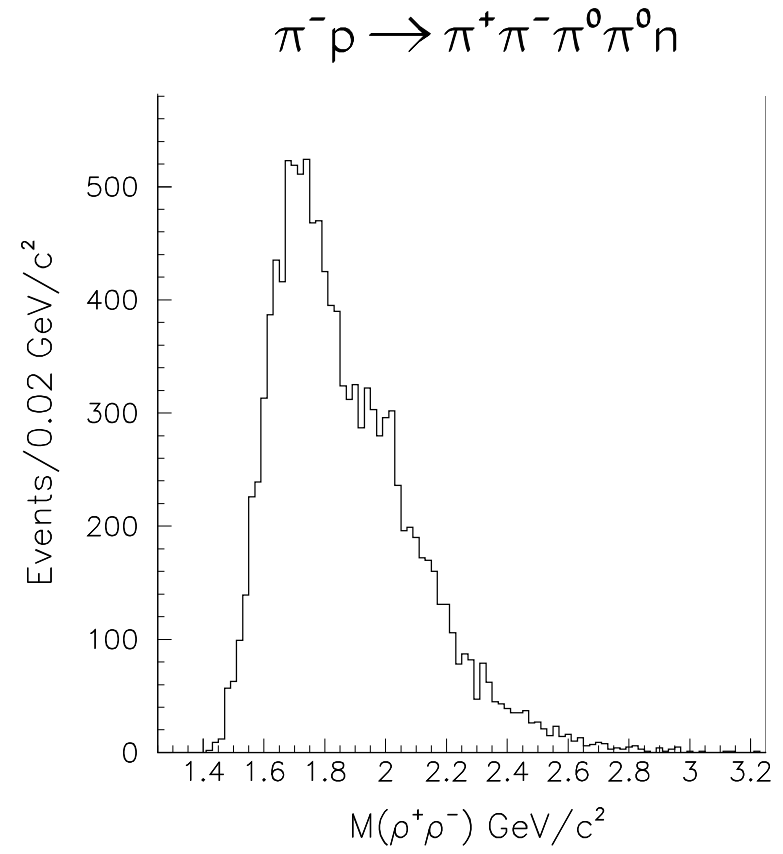

Figure 9: $\rho^+\rho^-$ mass plot.

# $\eta\pi^+\pi^-$

The $\eta\pi^+\pi^-$ final state gives access to both $a_0\pi$ and $a_2\pi$ via the $\eta\pi$ decay modes of the $a_0$ and the $a_2$ (15%). The well known $f_1(1285)$ has a substantial (44%) decay into $a_0\pi$. The decay $\eta' \to \eta\pi^+\pi^-$ (44%) is a fairly good signature in this channel. Figure 14 shows the strong $\eta'$ and $f_1(1285)$ peaks in $\eta\pi^+\pi^-$ and the $f_1(1285)$ and the $E/\iota(1420)$ in $a_0\pi$.

# $\eta\eta$

The $\eta\eta$ channel can be studied via both the $\eta \to \gamma\gamma$ and $\eta \to \pi^+\pi^-\pi^0$ modes. Figure 15 shows the $\eta\eta$ as measured in the 4 $\gamma$ mode. A similar response is found in the $\eta\eta$ channel in which one $\eta$ decays via $\pi^+\pi^-\pi^0$.

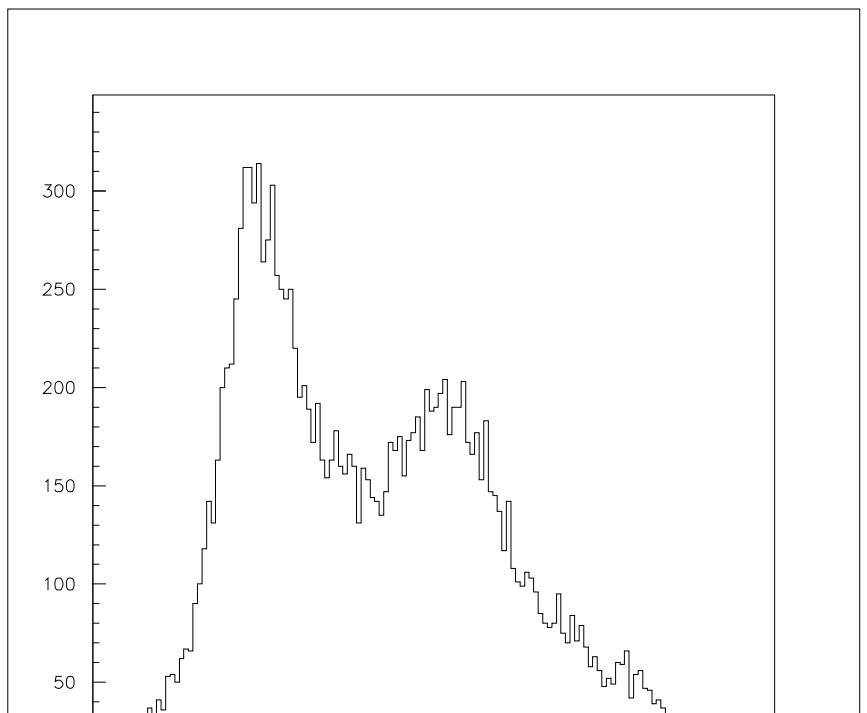

Figure 10: $\omega\pi^-$ mass plot. The $b_1$ and $\rho_3(1690)$ are clearly visible

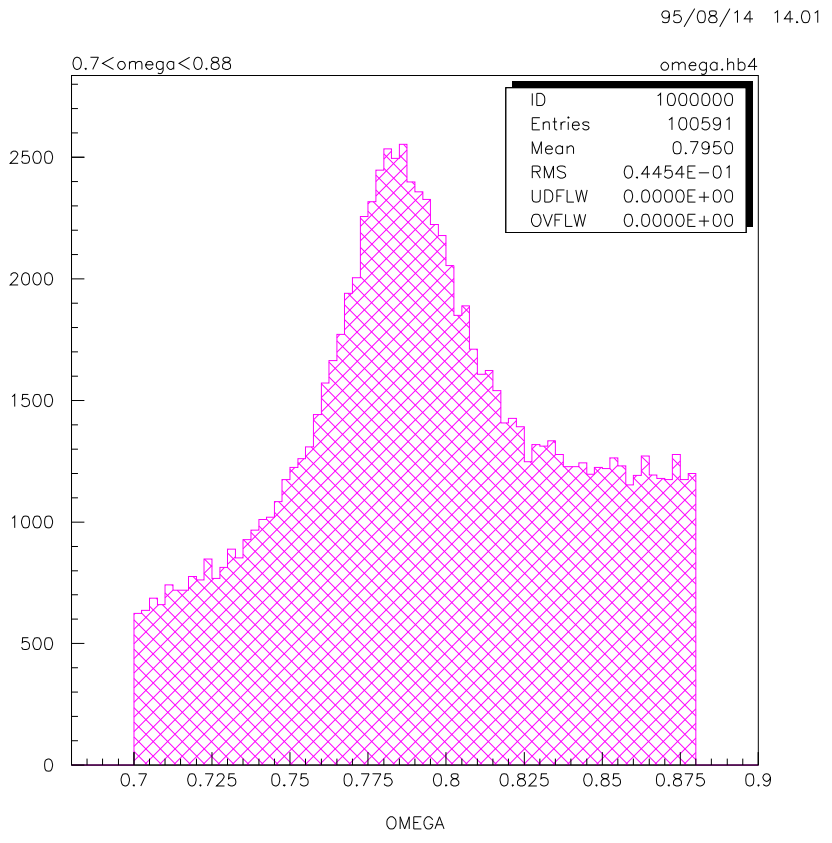

Figure 11: $\omega \to \pi^+\pi^0\pi^-$ in the 5 $\pi$ sample.

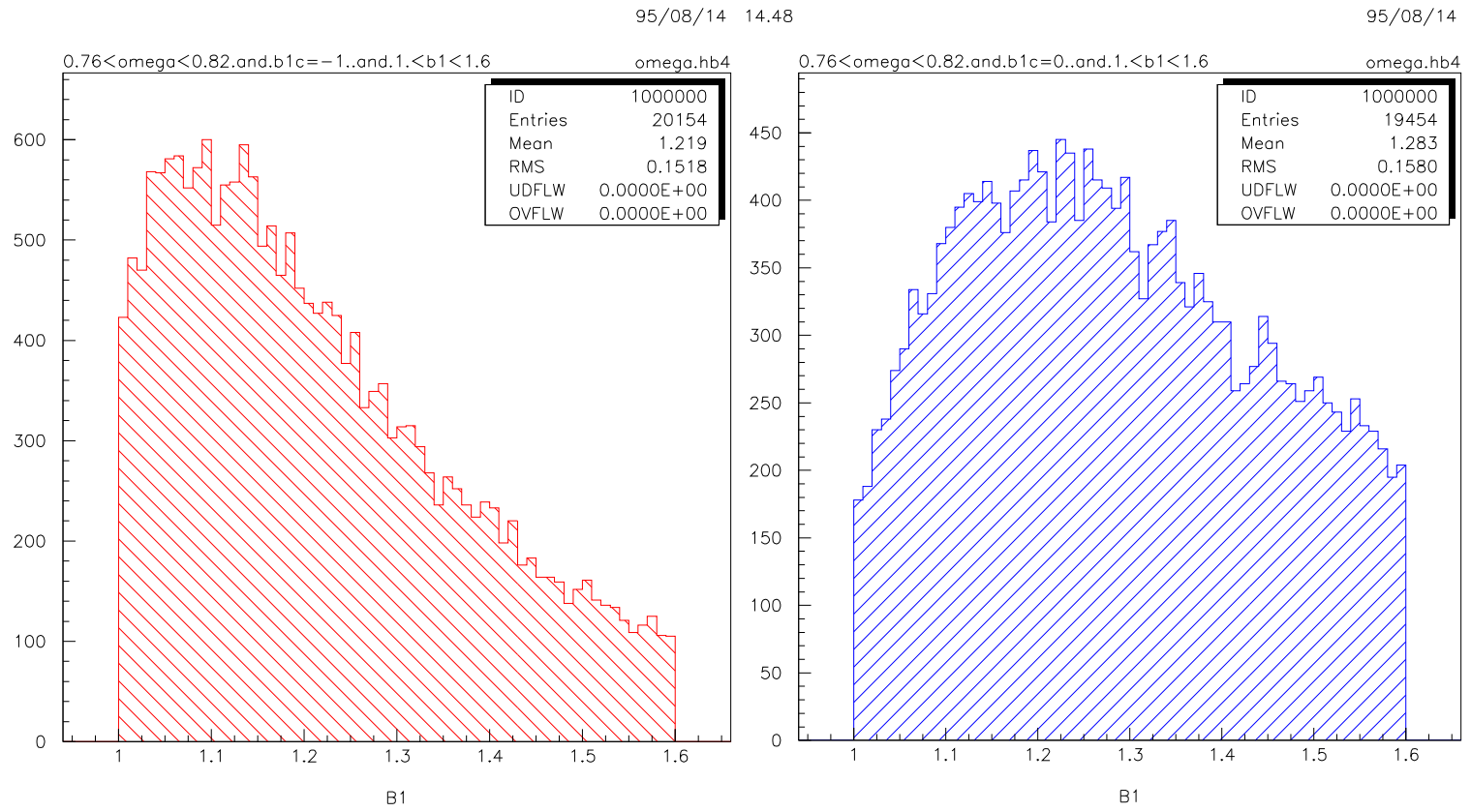

Figure 12: $\omega\pi^-$ and $\omega\pi^0$ mass plots in the 5 $\pi$ sample. There is no clear $b_1(1235) \to \omega\pi$ peak

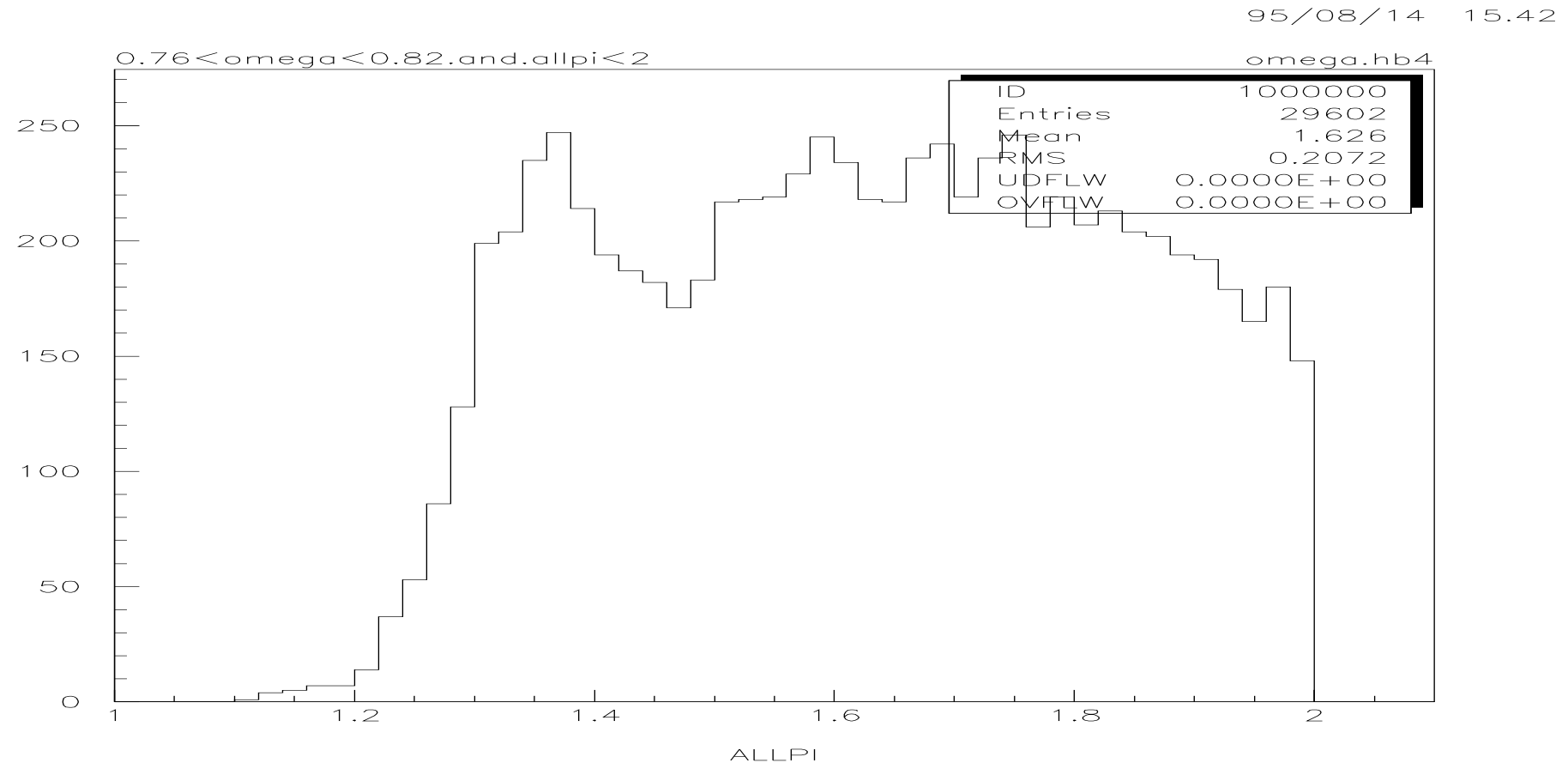

Figure 13: $\omega \to \pi^+\pi^0\pi^-$ in the 5 $\pi$ sample. Note the presence of the $a_2(1320)$

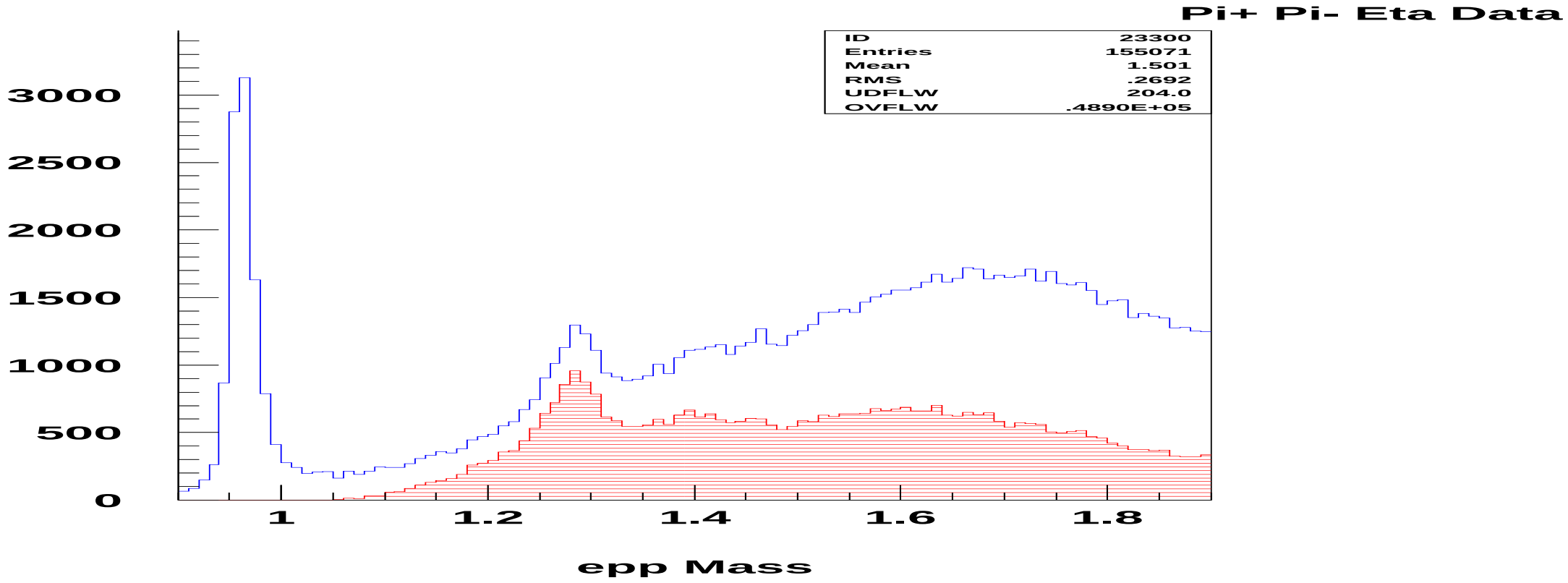

Figure 14: $\eta\pi^-\pi^+$ and $a_0\pi^0$ (stripped) mass plots. Note the clear signal for $\eta'(958)$ and the $f_1(1285)$.

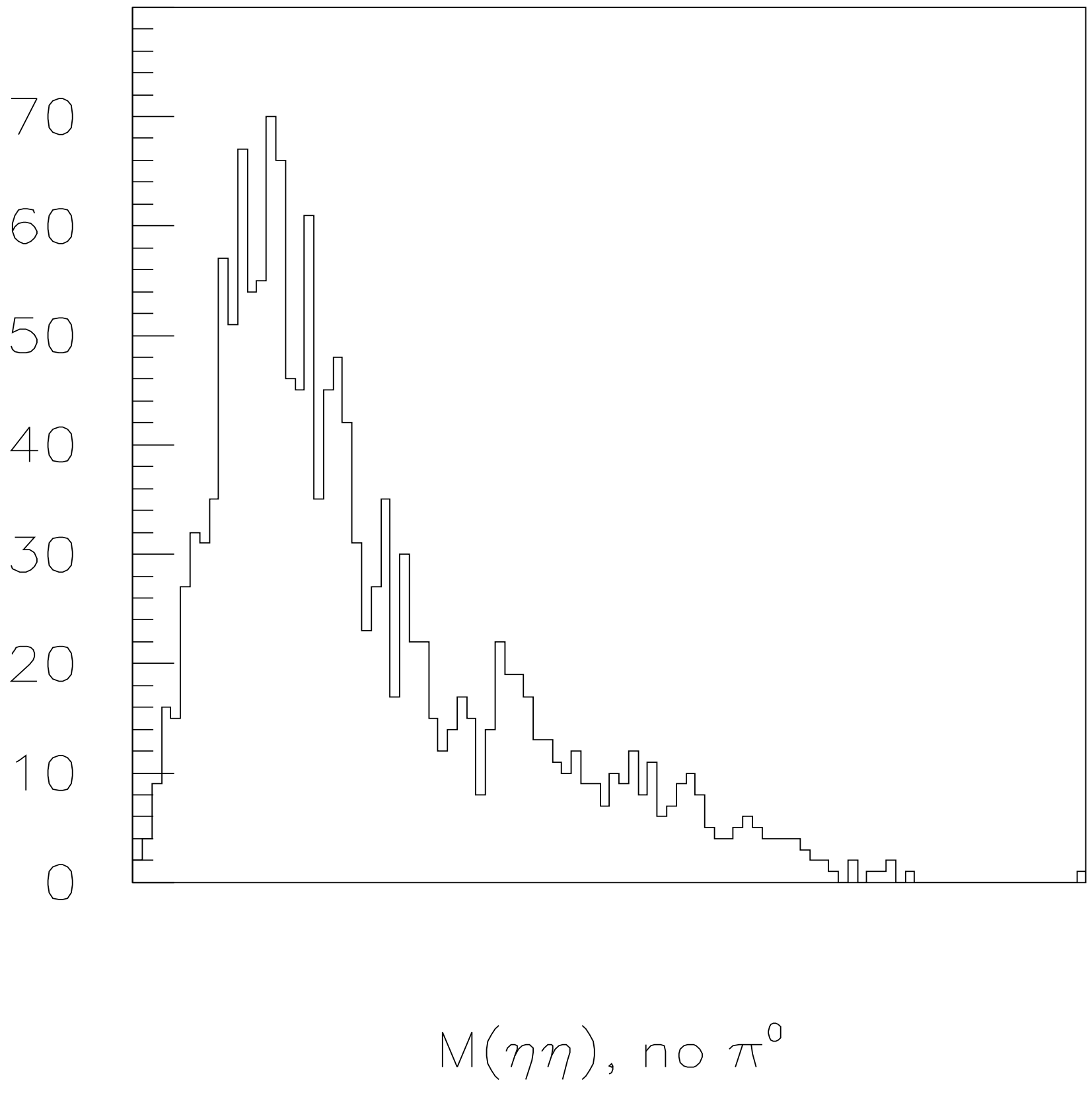

Figure 15: $\eta\eta \to 4\gamma$ mass plot.

# Conclusions

This experiment has access to a large number of final states that are favored in the decay of $q\bar{q}g$ exotics. Analysis of an initial sample of $2 \times 10^8$ events is progressing with some initial results presented here. An additional $8 \times 10^8$ events has been acquired in the 1995 run and reconstruction of those events is in progress.

For additional information on the data discussed here see the proceedings of the Hadron '95 meeting [5]. Results on many of the favored states can be expected during the next year.